\documentclass[preprint,12pt]{elsarticle}
\journal{International Journal of Plasticity}

\usepackage{graphics,epsfig}
\usepackage{graphicx}
\usepackage{float}
\usepackage{amssymb,amstext,amsmath}
\usepackage{mathtools}
\mathtoolsset{showonlyrefs}
\usepackage{booktabs}
\usepackage{natbib}
\usepackage[latin1]{inputenc}
\usepackage{epstopdf}

\begin{document}

\begin{frontmatter}
\title{Distribution of dislocations in twisted bars}
\author{K. C. Le\footnote{Corresponding author: phone: +49 234 32-26033, email: chau.le@rub.de.}, Y. Piao}
\address{Lehrstuhl f\"{u}r Mechanik - Materialtheorie, Ruhr-Universit\"{a}t Bochum,\\D-44780 Bochum, Germany}
\begin{abstract} 
An asymptotically exact continuum dislocation theory of single crystal bars under torsion is proposed. The dislocation distribution minimizing energy of the bar with zero torque is shown to be uniform. If the applied torque is non-zero, the minimizer exhibits a dislocation-free zone at the outer ring of the bar's cross-section. The non-uniform distribution of dislocations in equilibrium as well as the twist angle per unit length are found in terms of the given torque. With the energy dissipation being taken into account, there exists an elastic core region, while dislocation are concentrated in a ring between two dislocation-free zones. This leads to the change of the stress distribution increasing the critical threshold of the torque. 
\end{abstract}

\begin{keyword}
dislocations (A) \sep yield condition (A) \sep beams and columns (B) \sep asymptotic analysis  (C) \sep variational calculus (C).
\end{keyword}
\end{frontmatter}

\section{Introduction}

Dislocations appear to reduce energy of crystals. For large dislocation densities (which are typically in the range $10^8\div 10^{15}$ dislocations per square meter) it makes sense to use the continuum dislocation theory (CDT) to predict the distribution of dislocations in equilibrium. Although the framework of CDT has been laid down long time ago by \citet{Kondo52,Nye53,Bilby55,Kroener58,Berdichevsky67,Le-stumpf1996a,Le-stumpf1996b,Le-stumpf1996c}, the applicability of the theory became feasible only in recent years \citep{Berdichevsky06a,Gurtin2002,Ortiz99} thanks to the progress in averaging ensembles of randomly distributed dislocations \citep{Berdichevsky06b,Berdichevsky2016}. Among various alternative strain gradient plasticity theories we mention here only those of \citet{Fleck01,Gao99,Huang00,Huang04} which, in contrast to CDT, incorporate the history dependent plastic (or total) strain gradient rather than employ Nye-Bilby-Kr\"oner dislocation density tensor. 

In view of the variety of dislocation based gradient plasticity theories the necessity of having exact solutions of test problems, on which different models can be tested and compared, becomes obvious. One of the first investigated problem was the torsion of a single crystal bar. \citet{Fleck94} used strain gradient plasticity theory to compute the torque-twist curve and compared it with the experimental data showing clearly the size effect (see also \citep{Fleck01,Hwang03}). Based on an alternative strain gradient plasticity theory, \citet{Aifantis99} studied the same problem and found the size sensitivity of both yield stress and hardening rate. Later on, \citet{Kaluza2011} used CDT with the logarithmic energy of dislocation network proposed in \citep{Berdichevsky06a} to find not only the torque-twist curve, but also the distribution of dislocations in a twisted bar (see the application of CDT to other problems of dislocation pile-ups in a series of our papers \citep{Baitsch2015,Berdichevsky-Le07,Kochmann2008,Kochmann2009,Le2016,Le2013,Le2008a,Le2008b,Le2009}). More recently, \citet{Weinberger2011} computed the energy and distribution of screw dislocations in a free unloaded bar within the linear elasticity and provided the molecular dynamics simulations \citep{Weinberger2010} and dislocation dynamics simulations \citep{Akarapu2010,Espinosa2006} for the twisted bar. His results confirmed the qualitative agreement with our prediction in \citep{Kaluza2011} for the twisted bar but displayed at the same time some quantitative differences in the torque-twist curve and the distribution of dislocations in equilibrium.

Motivated by the above investigations as well as several existing discrepancies between theories and simulations, this paper studies the torsion of a single crystal bar with a constant circular cross-section using the asymptotically exact CDT proposed recently in \citep{Berdichevsky2016}. Our aim is to find the dislocation distribution in equilibrium as function of the given torque and compare with the similar results obtained by numerical simulations in \citep{Weinberger2011}. To simplify the analysis we assume that the crystal is elastically isotropic and all dislocations are screw. Besides, the side boundary of the bar is traction free and may therefore attract dislocations. We adopt the free energy formulation found by \citet{Berdichevsky2016} that is proved to be asymptotically exact in the continuum limit. If the dissipation can be neglected, the displacement and the plastic distortion should be found by the energy minimization. First, we show that the dislocation distribution minimizing energy of the bar with zero torque is uniform. This agrees well with the result obtained by \citet{Weinberger2011}. Next, for the bar loaded by a nonzero torque we find an energetic threshold for the dislocation nucleation. If the twist exceeds this threshold, excess dislocations appear to minimize the energy. It turns out that there is a dislocation-free zone at the outer ring of the bar's cross-section. The non-uniform distribution of dislocations in equilibrium as well as the twist angle are found in terms of the given torque. In case the dissipation due to the resistance to dislocation motion is taken into account, the energy minimization should be replaced by a variational equation. The solution is shown to have an elastic core region in the middle of the cross section. Dislocation are concentrated in a ring between two dislocation-free zones. This leads to the change of the stress distribution increasing the dissipative threshold of the torque. 

The paper is organized as follows. In Section 2 the setting of the problem is outlined and the energy minimizing dislocation distribution in a bar with zero torque is found. Section 3 studies the plastic torsion of the bar at zero dissipation through energy minimization. In Section 4, the plastic torsion of the bar at non-zero resistance to dislocation motion is analyzed. Finally, Section 5 concludes the present paper.

\section{Asymptotically exact energy of the bar}

\begin{figure}[htb]
	\centering
	\includegraphics[width=7cm]{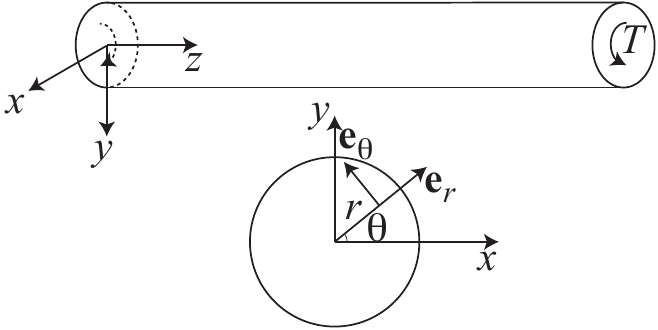}
	\caption{A single crystal bar loaded in torsion and its cross section}
	\label{fig:1}
\end{figure}

Consider a single crystal bar of length $L$ loaded in torsion by given torques $\pm T$ acting at its ends. Let $C$ be the cross section of the bar by planes $z=\text{const}$. For simplicity, we consider $C$ to be a circle of radius $R$ (see Fig.~\ref{fig:1}). The side boundary of the bar $\partial C\times [0,L]$ is free from tractions. The length of the bar is assumed to be much larger than the radius ($L\gg R$) to neglect the end effects. If the torque $T$ is sufficiently small, it is natural to assume that the bar deforms elastically so that the twist is proportional to the torque, provided the bar is initially dislocation-free. If $T$ exceeds some critical value, then screw dislocations may appear. We assume that the active slip planes are perpendicular to the vectors $\textbf{e}_\theta $ in the cylindrical coordinate system, while the slip directions as well as the dislocation lines are parallel to the $z$-axis.  Mention that this assumption is not likely to be realistic for single crystals. However, since the Burgers vector is parallel to a screw dislocation, any crystallographic plane containing the dislocation is a possible slip plane. Thus, the screw dislocations in single cubic primitive crystals with the Burgers vector parallel to $[001]$ and slip planes of type $\{ mn0\}$, $m,n$ being any irreducible pair of integers, may approximately be regarded as representing this situation. Likewise, the screw dislocations in single hcp-crystals with the Burgers vector parallel to $[0001]$ and slip planes of type $\{ 11\bar{2}0\}$ may also be considered as satisfying this assumption to some degree. Our aim is to determine the distribution of dislocations as function of $T$ by the continuum dislocation theory. For screw dislocations with the slip planes perpendicular to the vector $\textbf{e}_\theta $, the tensor of plastic distortion, $\beta _{ij}$, has only one non-zero component $\beta _{z\theta }\equiv \beta $. Function $\beta $ can be interpreted as the plastic warping, and, by the symmetry reasoning, we assume that $\beta $ depends only on $r$-coordinate: $\beta =\beta (r)$. The only non-zero components of the plastic strain tensor are given by
\begin{equation*}
\varepsilon ^p_{\theta z}=\varepsilon ^p_{z\theta }=\frac{1}{2}\beta
(r).
\end{equation*}
For the bar of circular cross-section loaded in torsion the only non-zero components of the displacement vector is $u_\theta $ which depends only on $r$ and $z$: $u_\theta =u_\theta (r,z)$. With the previous equation we obtain for the components of the elastic strain tensor
\begin{equation}\label{2.2}
\begin{split}
\varepsilon ^e_{rr}=\varepsilon ^e_{\theta \theta}=\varepsilon ^e_{rz}=\varepsilon ^e_{zz}=0,
\\
\varepsilon ^e_{r\theta }=\frac{1}{2}(u_{\theta ,r}-u_\theta /r),\quad \varepsilon ^e_{\theta z}=\frac{1}{2}(u_{\theta ,z}-\beta ),
\end{split}
\end{equation}
where the comma in indices denotes the partial derivative with respect to the corresponding coordinate. The only non-zero component of the tensor of dislocation density, $\alpha _{ij}=\varepsilon _{jkl}\beta _{il,k}$, is
\begin{equation*}
\alpha _{zz}=\beta _{,r}+\beta /r.
\end{equation*}
Mention that $\alpha _{zz}da$ presents the resultant Burgers vector of all dislocations, whose dislocation lines cross an infinitesimal area $da$ perpendicular to the $z$-axis. Thus, the number of screw dislocations per unit area becomes 
\begin{equation}\label{2.2a}
\rho =\frac{1}{b}|\alpha _{zz}|=\frac{1}{b}|\beta _{,r}+\beta /r|,	
\end{equation}
with $b$ being the magnitude of Burgers vector. We require the dislocation density to remain finite everywhere including $r=0$, so function $\beta (r)$ must satisfy the regularity condition
\begin{equation}\label{2.2b}
\beta (0)=0.	
\end{equation}
Hence, the center line of the bar can be considered as an obstacle hindering the motion of dislocations which have to pile up against it.

Following \citet{Kroener92} we regard the elastic strain $\varepsilon ^e_{ij}$ and the dislocation density $\alpha _{ij}$ as the state variables of the continuum dislocation theory. The free energy per unit volume of the crystal with dislocations is allowed to depend only on the position vector $\mathbf{x}=(x,y,z)$ as well as on these state variables. In the continuum limit we lay down
\begin{equation}
\phi (\mathbf{x},\varepsilon ^e_{ij},\alpha _{ij})=\phi _0(\varepsilon ^e_{ij})+\phi _m(\mathbf{x},\alpha _{ij}). \label{2.3}
\end{equation}
The first term of \eqref{2.3} corresponds to the energy contribution due to the elastic strain that includes also the energy of interaction of dislocations, while the second term describes the self-energy of dislocations whose explicit dependence on $\mathbf{x}$ accounts for the influence of the free boundary \citep{Eshelby1953}. \citet{Berdichevsky2016} found the free energy density of  elastically isotropic crystal bar containing screw dislocations which we present in the form
\begin{equation}
\label{2.4}
\phi (\mathbf{x},\varepsilon ^e_{ij},\alpha _{ij})=\frac{1}{2}\mu (\varepsilon_{r\theta }^{e})^2+ \frac{1}{2}\mu (\varepsilon_{\theta z}^{e})^2+\frac{\mu b^2}{4\pi } f(r/R) \rho ,
\end{equation}
where $\mu$ is the shear modulus, $f(r/R)=\ln (1-r^2/R^2)+\ln \frac{R}{r_0}+1/4$, with $r_0$ being the cut-off radius of the dislocation core. The first two terms of \eqref{2.4} correspond to the energy contribution due to the elastic strain, the third term is the self-energy of dislocations including dislocation-boundary interaction energy. Note that function $f(r/R)$ is well defined only for $r$ having a distance larger than $r_0/2$ to the side boundary $r=R$. For $r\in (R-r_0/2,R)$ we set $f(r/R)=1/4$. As shown in \citep{Berdichevsky2016}, this energy density is asymptotic exact in the continuum limit, when the number of dislocations becomes large. According to Berdichevsky's classification, the first two terms are the main terms, while the last term belongs to the small correction term that is comparable with the error in obtaining the first two terms in the continuum limit for the ensemble of randomly distributed dislocations and should therefore be neglected. However, if we do averaging in the spatial way in which the volume of the bar is divided into a large number of boxes such that the number of dislocations in each box is proportional to the dislocation density times the box volume, the errors in obtaining all three terms through averaging have the same order of smallness that tends to zero as sizes of the boxes go to zero. Thus, the last term in the average energy density \eqref{2.4}, although small compared to the first two due to the smallness of $b$, has the right of existence in the continuum theory obtained by the spatial averaging procedure. We will see that this small correction term enables one to uniquely determine the distribution of dislocations in terms of the torque. 

\begin{figure}[htb]
	\centering
	\includegraphics{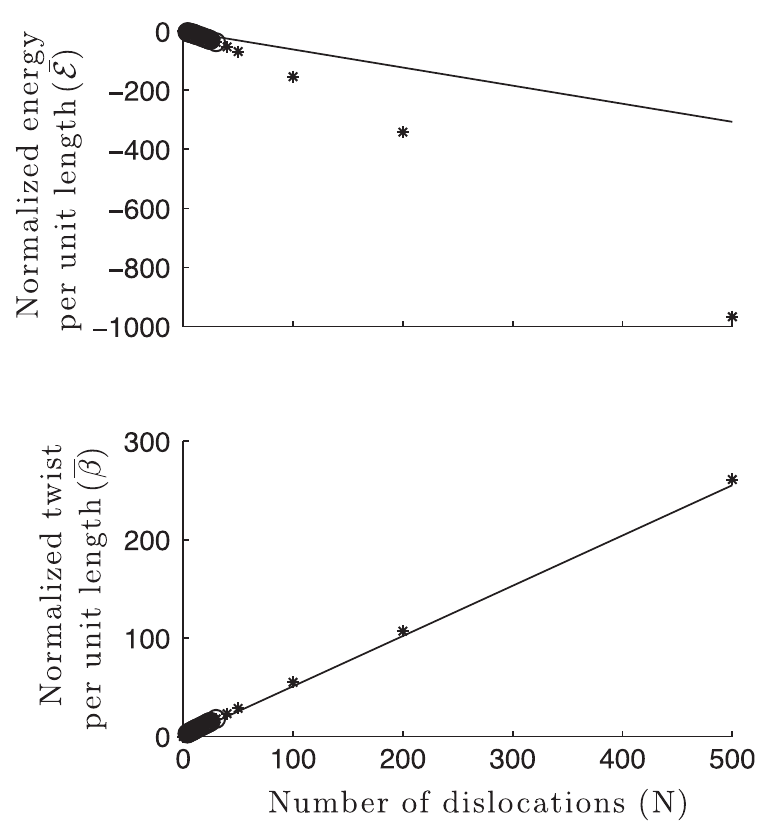}
	\caption{The normalized energy and normalized twist per unit length as functions of the number of dislocations: (i) bold line: CDT, (ii) points: numerical simulations \citep{Weinberger2011}.}
	\label{fig:2}
\end{figure}

We assume that the distributed tractions $t_\theta =\pm t(r)$ leading to the torques $\pm T$ act at the ends $z=0,L$ of the bar. Under the assumption of axial symmetry the energy functional of twisted bar reads
\begin{multline}
I(u_\theta ,\beta)=2\pi \int_0^L \int_{0}^{R} \biggl[ \frac{1}{2}\mu ( u_{\theta ,r}-\frac{1}{r}u_\theta )^2+\frac{1}{2}\mu (u_{\theta ,z}-\beta )^2 
\\
+ \frac{\mu b}{4\pi } f(r/R) |\beta _{,r}+\beta /r|
\biggr] r dr dz - 2\pi \int_{0}^{R} t(r)[u_\theta (r,L)-u(r,0)]rdr . \label{2.5}
\end{multline}
We consider first the case of free bar with zero traction $t(r)=0$ causing no torque. In this case the true displacement and plastic warping minimize the energy functional
\begin{multline}
\label{2.6}
I(u_\theta ,\beta)=2\pi \int_0^L \int_{0}^{R} \biggl[ \frac{1}{2}\mu ( u_{\theta ,r}-\frac{1}{r}u_\theta )^2+\frac{1}{2}\mu (u_{\theta ,z}-\beta )^2 
\\
+ \frac{\mu b}{4\pi } f(r/R) |\beta _{,r}+\beta /r|
\biggr] r dr dz  
\end{multline}
among all admissible $u_\theta $ and $\beta $. Since functional \eqref{2.6} is non-negative, the minimizer vanishes identically: $u_\theta =0$ and $\beta =0$. There is another interesting question in the case of zero torque, first raised by \citet{Eshelby1953} and studied later in \citep{Weinberger2011}: provided $N$ screw dislocations exist in the free unloaded bar, how to find the energy minimizing dislocation distribution. Within the CDT proposed in this paper, the problem reduces to finding $u_\theta $ and $\beta $ that minimize \eqref{2.6} under the constraint
\begin{equation}
\label{2.7}
2\pi \int_0^R \rho rdr=\frac{2\pi}{b}\int_0^R |\beta _{,r}+\beta /r| rdr=N.
\end{equation}
This problem can be solved by the variational-asymptotic method (see, e.g., \citep{Le1999}). Since the last term in functional \eqref{2.6} is small, we neglect it in the first approximation. Then the minimizer that makes two first terms vanishing reads
\begin{equation}\label{2.7a}
u_\theta =\omega _0rz, \quad \beta =\omega _0r,
\end{equation}
where $\omega _0$ can be interpreted as the twist per unit length due to screw dislocations in the free bar. This leads to the uniform dislocation distribution $\rho =2\omega _0/b$. Assuming that $\omega _0>0$ and substituting $\rho $ into \eqref{2.7}, we obtain
\begin{displaymath}
2\pi \int_0^R 2\omega _0/b rdr=\frac{2\omega _0}{b}\pi R^2=N \quad \Rightarrow \quad \omega _0=\frac{Nb}{2\pi R^2}.
\end{displaymath}
If we denote the dimensionless twist per unit length as $\bar{\omega }_0=\omega _0/(b/\pi R^2)$, then $\bar{\omega }_0=N/2$. This agrees well with the result obtained by \citet{Weinberger2011} for large $N$ as shown in Fig.~\ref{fig:2}.\footnote{In Weinberger's paper the normalized twist per unit length was denoted by $\bar{\beta }$.} The minimum of energy is found by substituting \eqref{2.7a} into the energy functional \eqref{2.6}. This yields the energy per unit length
\begin{displaymath}
\mathcal{E}=2\pi  \frac{\mu b^2}{4\pi } \int_0^R 2\omega _0/b f(r/R)rdr=\frac{\mu b^2}{2\pi}(\frac{N}{2}\epsilon _0+\bar{\mathcal{E}}),
\end{displaymath}
where $\epsilon _0=\ln \frac{R}{r_0}+1/4$ and
\begin{displaymath}
\bar{\mathcal{E}}=N\int_0^1\ln (1-\xi ^2)\xi d\xi =-\frac{N}{2}.
\end{displaymath}
The comparison with the dimensionless energy $\bar{\mathcal{E}}$ obtained by \citet{Weinberger2011} is shown in the same Figure. In contrast to the twist, there is a difference in the slope of $\bar{\mathcal{E}}$: in our continuum theory this slope is $-1/2$, while Weinberger's numerical simulations gives $-2$ for large $N$. Since functional \eqref{2.6} is asymptotically exact for large $N$ \citep{Berdichevsky2016}, Weinberger's numerical calculations for $\bar{\mathcal{E}}$ must contain some error. 

Assume now that a non-zero torque $T$ is applied. In this case the true displacement and plastic warping should be found from minimizing functional \eqref{2.5}. This variational problem can again be solved by the variational-asymptotic method. In the first step we keep in \eqref{2.5} only the first term as the main asymptotic term \citep{Le1999}. Then the true displacement $u_\theta $ reads
\begin{equation*}
u_\theta (r,z)=\omega rz,
\end{equation*}
where $\omega =\varphi /L$ corresponds to the twist of the bar per unit length. Except the two edges of the bar where $u_\theta $ and $\beta $ may depend on the detailed distribution of $t(r)$,  the energy per unit length in the main part of the bar reduces then to
\begin{equation}
\bar{I}(\beta)=2\pi \int_{0}^{R} \biggl[ \frac{1}{2}\mu ( r\omega -\beta )^2  
+  \frac{\mu b}{4\pi } f(r/R) |\beta _{,r}+\beta /r|
\biggr] r dr. \label{2.8}
\end{equation}
If the resistance to dislocation motion is negligible (and, hence, the dissipation is zero), the true plastic warping minimizes functional \eqref{2.8} among all admissible function $\beta (x)$ satisfying the regularity condition \eqref{2.2b}. 

If the resistance to dislocation motion cannot be neglected, then the energy minimization must be replaced by the variational equation \citep{Sedov65,Berdichevsky67}
\begin{equation}
\delta \bar{I}+2\pi \int_0^R \frac{\partial D}{\partial \dot{\beta }}\delta \beta rdr=0, \label{2.9}
\end{equation}
where, in case of rate-independent theory, 
\begin{equation*}
D(\dot{\beta})=K|\dot{\beta}| .
\end{equation*}
Function $D(\dot{\beta})$ represents the dissipation potential due to plastic warping, with $K$ being called critical resolved shear stress, and the dot above a function denoting its time derivative. For $\dot{\beta }=0$ the variational equation \eqref{2.9} needs not be satisfied: it is replaced by the equation $\dot{\beta }=0$.

\section{Dislocation distribution at zero dissipation}\label{sec:nuc}
We first analyze the situation when the resistance to dislocation motion is negligible (and, hence, the dissipation is zero). In this case the determination of $\beta (r)$ reduces to the minimization problem \eqref{2.8}. It is convenient to introduce the following dimensionless variable and quantities
\begin{equation}\label{2.10}
\xi =\frac{r}{R}, \quad \kappa=R\omega , \quad E=\frac{\bar{I}}{2\pi \mu R^2}, \quad \chi =\frac{b}{4\pi R}.
\end{equation}
The dimensionless variable $\xi $ changes on the interval $(0,1)$. Assuming for definiteness that the dimensionless dislocation density $\beta_{,\xi }+\beta /\xi $ is positive, functional \eqref{2.4} reduces to
\begin{equation}\label{2.11}
E(\beta )=\int_0^1 [\frac{1}{2}(\kappa \xi -\beta )^2+\chi f(\xi ) (\beta_{,\xi }+\beta /\xi ) ]\, \xi d\xi ,
\end{equation}
where $f(\xi )=\ln (1-\xi ^2)+\epsilon _0$. We minimize functional \eqref{2.11} among functions $\beta (r)$ satisfying the regularity condition $\beta(0)=0$. Because the last gradient term is linear in $\beta _{,\xi }$, we allow the plastic distortion to have jump at some point $\xi =l$.

Under the action of Peach-Koehler force (see, e.g., \cite{Le2010}) dislocations move toward the middle line of the bar, so there should be a dislocation-free zone near the side boundary: $\beta _{,\xi }+\beta /\xi =0$. The latter equation is satisfied only if $\beta =\beta _0/\xi $ for some constant $\beta _0$. This leads to the following Ansatz for the minimizer
\begin{equation}\label{2.12}
\beta (\xi )=
  \begin{cases}
    \beta _1(\xi ) & \text{for $\xi \in (0,l)$}, \\
    \beta _0/\xi & \text{for $\xi \in (l,1)$}, 
  \end{cases}
\end{equation}
where $l$ is an unknown length, $0\le l\le 1$. We admit that $\beta (\xi )$ may have a jump at $\xi =l$. We have to find $\beta _1(\xi )$ and the constants, $\beta _0$ and $l$. Thus, the functional becomes
\begin{equation}\label{2.13}
E=\int_0^l[\frac{1}{2}(\kappa \xi -\beta _1)^2+\chi f(\xi ) (\beta _{1,\xi }+\frac{\beta _1}{\xi })
 ]\, \xi d\xi +\int_l^1\frac{1}{2}(\kappa \xi -\frac{\beta _0}{\xi})^2\, \xi d\xi .
\end{equation}
Varying energy functional \eqref{2.13} with respect to $\beta _1(\xi)$ we obtain the (non-differential) equation for it on the interval $(0,l)$
\begin{equation}\label{2.14}
-(\kappa \xi -\beta _1)\xi -\chi f^\prime (\xi )\xi =0,
\end{equation}
yielding
\begin{equation}
\label{2.15}
\beta _1(\xi )=\kappa \xi +\chi f^\prime (\xi ).
\end{equation}
Due to the specific linear dependence of energy functional on $\beta _{1,\xi }$ leading to the (non-differential) equation for $\beta _1(\xi)$, $\beta _1(l)$ cannot be varied arbitrarily, so $\delta \beta _1(l)=0$. However, the variation of \eqref{2.13} with respect to $l$ and $\beta _0$ yields two additional conditions 
\begin{equation}\label{2.16}
\begin{split}
\frac{1}{2}(\kappa l-\beta _1(l))^2+\chi f(l) (\beta _{1,\xi }+\frac{\beta _1}{l})- \frac{1}{2}(\kappa l -\frac{\beta _0}{l})^2=0, 
\\
\int_l^1 (\kappa \xi -\frac{\beta _0}{\xi }) d\xi =0.
\end{split}
\end{equation}

From \eqref{2.16}$_2$ we find that
\begin{equation}
\label{2.16a}
\beta _0=-\frac{\kappa (1-l^2)}{2\ln l}.
\end{equation}
Because $l<1$, $\beta _0>0$ if $\kappa >0$. Plugging \eqref{2.15} and \eqref{2.16a} into \eqref{2.16}$_1$, we obtain the transcendental equation to determine $l$
\begin{equation}\label{2.16b}
\frac{1}{2}\chi ^2(f^{\prime }(l))^2+2\kappa \chi f(l)+\chi ^2f(l)\left[ f^{\prime \prime }(l)+\frac{f^\prime (l)}{l}\right] -\frac{1}{2}\kappa ^2\left( l+\frac{1-l^2}{2l \ln l}\right) ^2 =0.
\end{equation}
We may either solve this equation with respect to $l$ in terms of $\kappa $, or use $l$ as the parameter and find $\kappa $ in terms of $l$. With respect to $\kappa $, equation \eqref{2.16b} can be presented as the quadratic equation
\begin{equation}\label{2.16c}
a(l)\kappa ^2-2b(l)\kappa -c(l)=0,
\end{equation}
where
\begin{gather*}
a(l)=\left( l+\frac{1-l^2}{2l \ln l}\right) ^2, \quad b(l)=2\chi f(l),
\\
c(l)=\chi ^2(f^{\prime }(l))^2+2\chi ^2f(l)\left[ f^{\prime \prime }(l)+\frac{f^\prime (l)}{l}\right].
\end{gather*}
Provided the discriminant $b^2(l)+a(l)c(l)$ is positive, we take the positive root of \eqref{2.16c} yielding
\begin{equation}\label{2.17}
\kappa (l)=\frac{b(l)+\sqrt{b^2(l)+a(l)c(l)}}{a(l)}.
\end{equation}
The smallest $l=l_m$ that gives a real double root $\kappa _m=b(l_m)/a(l_m)$ is found from the equation
\begin{displaymath}
b^2(l_m)+a(l_m)c(l_m)=0.
\end{displaymath}
As will be seen later, this $l_m$ indicates the onset of dislocation nucleation. 

Having $\beta (\xi )$ according to \eqref{2.12}, \eqref{2.15}, \eqref{2.16a}, we find the signed dislocation density from \eqref{2.2a}. For $\xi \in (0,l)$ we have
\begin{equation}
\label{2.17a}
\rho =\frac{1}{b}(\beta _{,r}+\frac{\beta }{r})=\frac{1}{bR}(\beta _{,\xi }+\frac{\beta }{\xi})=\frac{1}{bR}\left[ 2\kappa +\chi (f^{\prime \prime }(\xi)+\frac{f^\prime (\xi)}{\xi}) \right] .
\end{equation}
For $\xi \in (l,1)$ we know that $\rho =0$. With this dislocation density we can also compute the total number of dislocations in a bar 
\begin{equation}
\label{2.17b}
N=2\pi \int_0^R \rho rdr=2\pi \frac{R}{b}(\kappa l^2-\chi \frac{2l^2}{1-l^2}).
\end{equation}
The dimensionless shear stress distribution $\tau (\xi )=\sigma _{\theta z}(\xi )/\mu $ is given by
\begin{equation}
\label{2.17c}
\tau (\xi )=\begin{cases}
 -\chi f^\prime (\xi )   & \text{for $\xi \in (0,l)$}, \\
  \kappa \xi + \frac{\kappa (1-l^2)}{2\ln l \, \xi }  & \text{for $\xi \in (l,1)$}.
\end{cases}
\end{equation}
Note that the jump in shear stress across the cylindrical surface $\xi =l$ does not violate the equilibrium condition. The torque is computed as the resultant moment of this shear stress or, alternatively, is obtained by differentiating functional \eqref{2.11} with respect to $\kappa $. This gives
\begin{equation*}
\bar{T}=\frac{\partial E}{\partial \kappa}=\frac{T}{2\pi \mu R^3}=\int_0^1 (\kappa \xi -\beta )\xi ^2d\xi .
\end{equation*}
To compute the dimensionless torque $\bar{T}$ we substitute $\beta (\xi )$ from \eqref{2.12} into this integral. Using the above solution for $\beta (\xi )$ we easily find that
\begin{equation}
\label{2.18}
\bar{T}=-\chi (l^2+\ln (1-l^2))+\frac{\kappa }{4}(1-l^4)+\frac{\kappa (1-l^2)^2}{4\ln l}.
\end{equation} 
For the twist $\kappa <\kappa _m$ the plastic warping $\beta (\xi )$ must be identically zero, so we have the purely elastic solution with $u_\theta (\xi )=\kappa \xi $ and $\beta (\xi )=0$. In this case the torque $\bar{T}$ is proportional to the twist $\kappa $: 
\begin{displaymath}
\bar{T}=\frac{\kappa }{4}.
\end{displaymath}

\begin{figure}[htb]
	\centering
	\includegraphics[width=8cm]{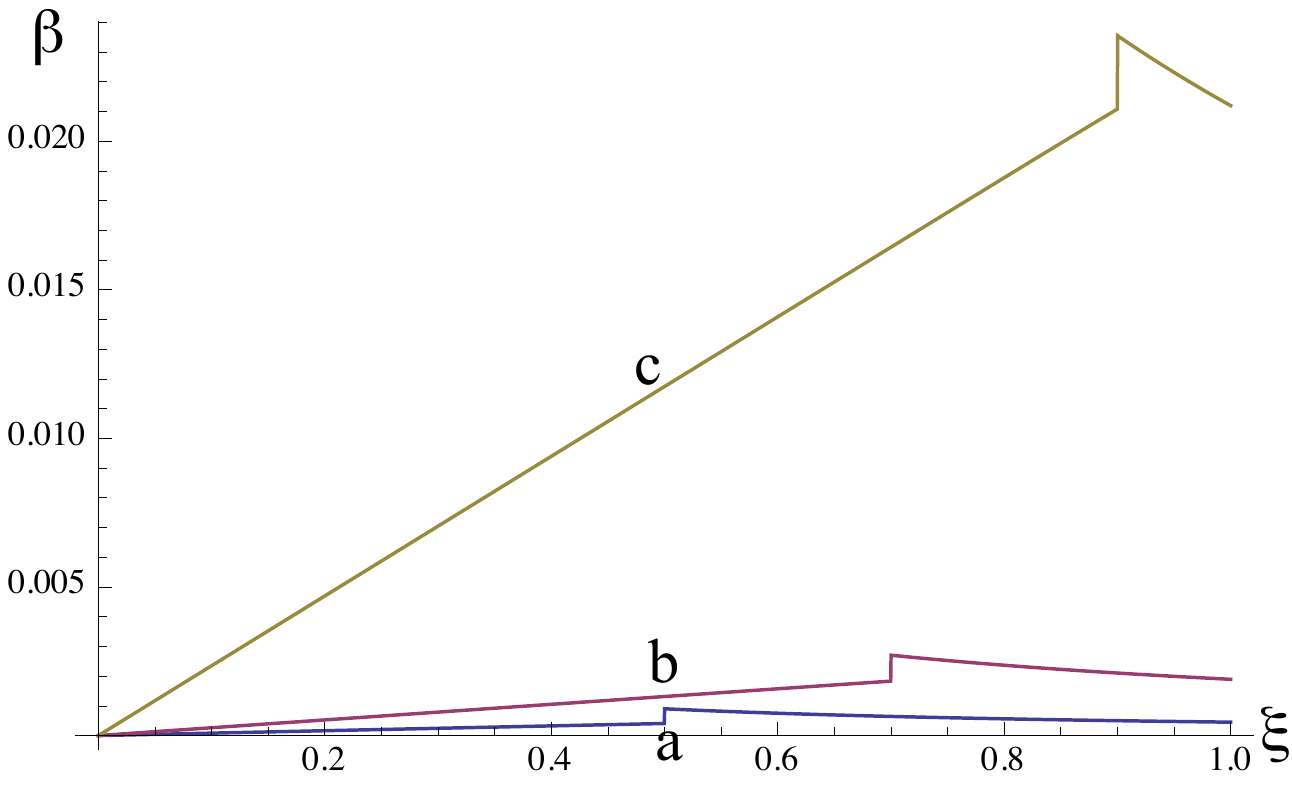}
	\caption{Evolution of the plastic warping $\beta (\xi )$: (a) $l=0.5$ ($\kappa =0.00083$), (b) $l=0.7$ ($\kappa =0.00265$), (c) $l=0.9$ ($\kappa =0.02351$).}
	\label{fig:3}
\end{figure}

Fig.~\ref{fig:3} shows the evolution of $\beta (\xi )$ as $\kappa $ increases. For the simulation we took $R=1$micron, $b=r_0=1$\AA . We see that $\beta (\xi )$ increases as $\kappa $ increases. Besides, the plastic warping exhibits a jump at $\xi =l$ that also increases as $\kappa $ increases. Since the total strain is continuous, such jump indicates the misorientation of the lattice across the surface $\xi =l$.

\begin{figure}[htb]
	\centering
	\includegraphics[width=8cm]{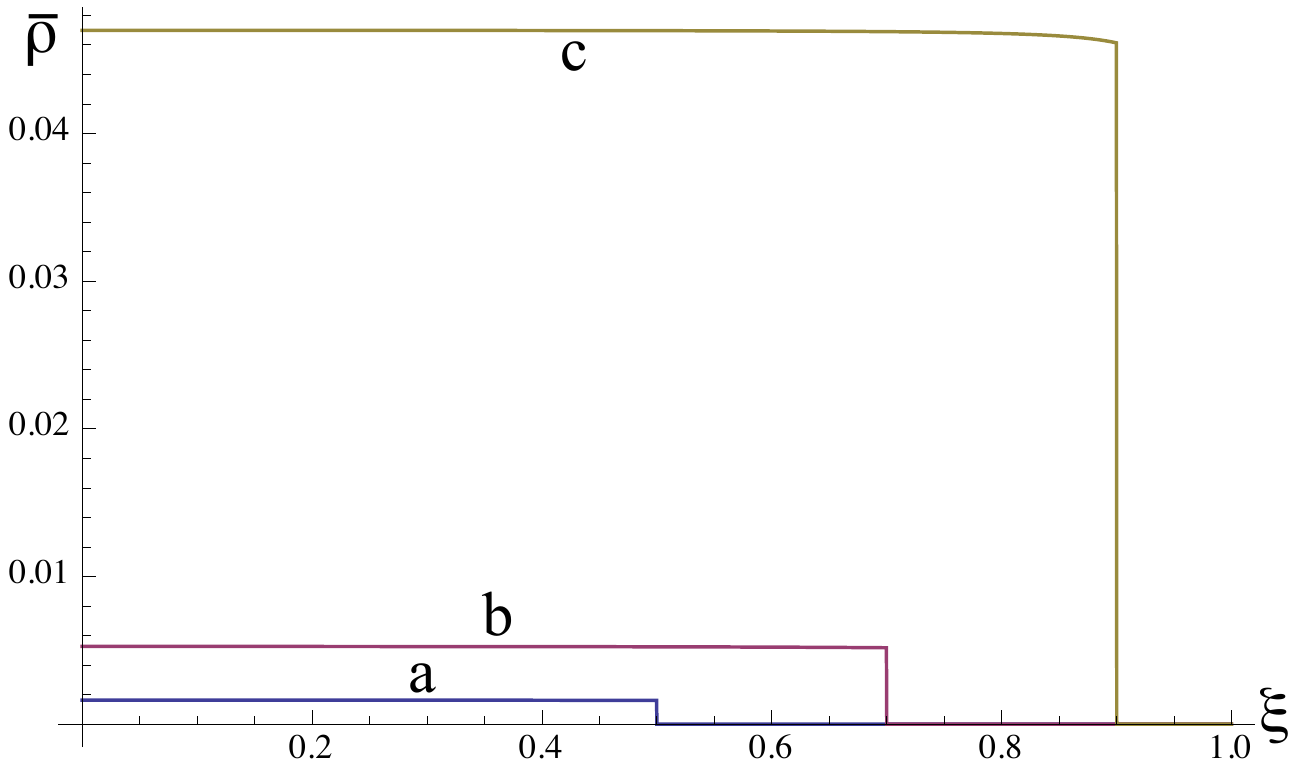}
	\caption{Distribution of normalized dislocation density $\bar{\rho }(\xi )$: (a) $l=0.5$ ($\kappa =0.00083$), (b) $l=0.7$ ($\kappa =0.00265$), (c) $l=0.9$ ($\kappa =0.02351$).}
	\label{fig:4}
\end{figure}

Fig.~\ref{fig:4} shows the distributions of the normalized dislocation density $\bar{\rho }(\xi )=bR \rho (\xi )$ as $\kappa $ increases. One can see that the dislocation-free zone diminishes as $\kappa $ increases. For small $\kappa $ the dislocation density remains nearly constant. For large $\kappa $ we see some influence of the free boundary on the distribution of dislocations: the density is slightly decreases with $\xi $. 

\begin{figure}[htb]
	\centering
	\includegraphics[width=8cm]{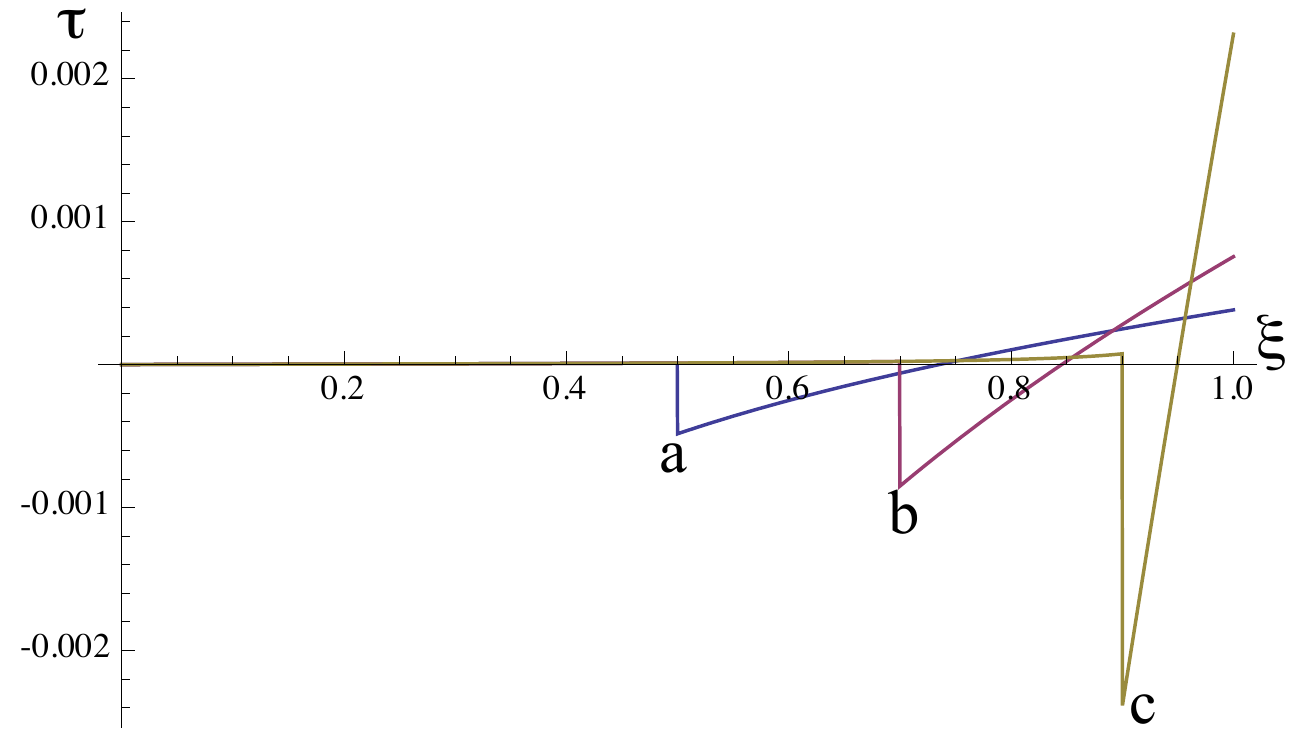}
	\caption{Distribution of dimensionless shear stress $\tau (\xi )$: (a) $l=0.5$ ($\kappa =0.00083$), (b) $l=0.7$ ($\kappa =0.00265$), (c) $l=0.9$ ($\kappa =0.02351$).}
	\label{fig:4a}
\end{figure}

\begin{figure}[htb]
	\centering
	\includegraphics[width=12cm]{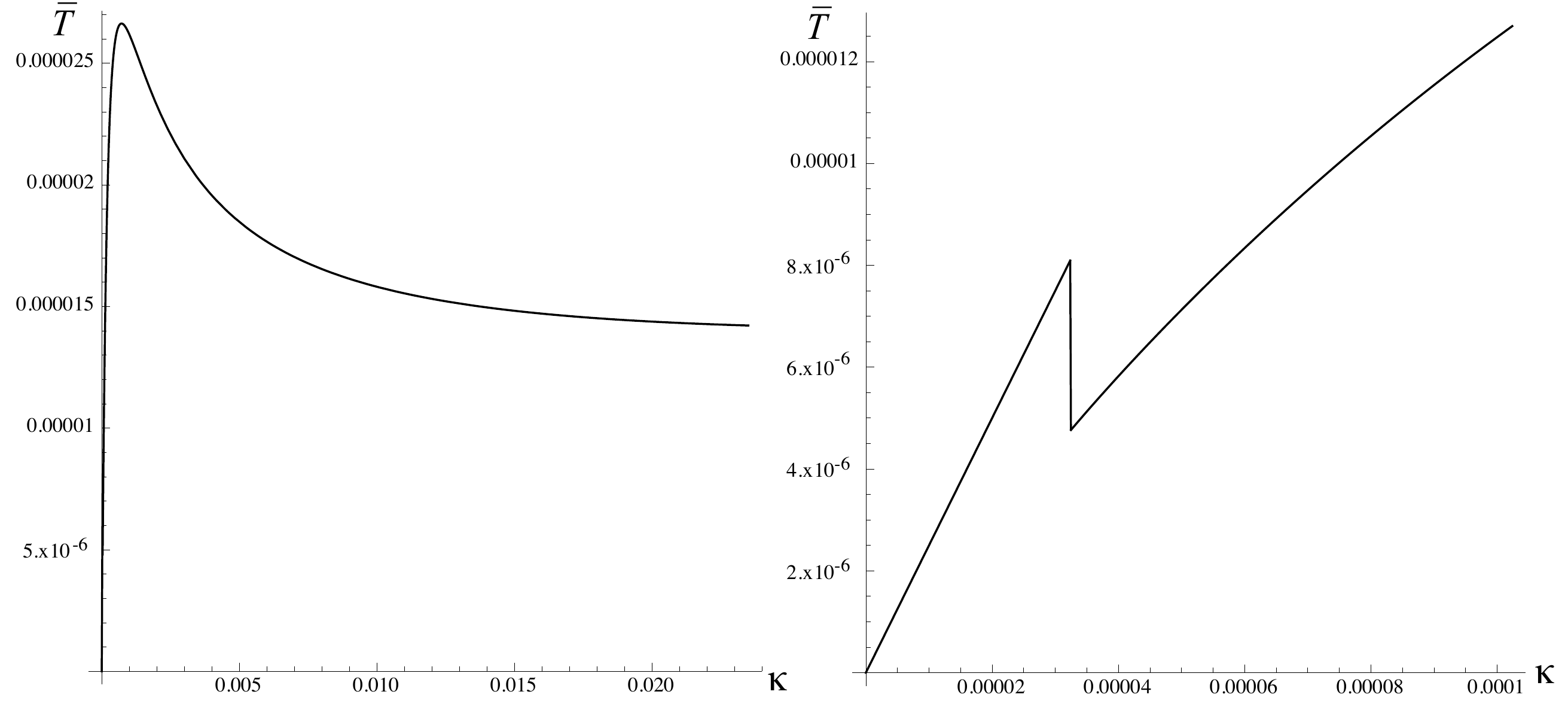}
	\caption{The torque-twist curve: (i) Left: the whole curve, (ii) Right: Zoomed near the origin.}
	\label{fig:5}
\end{figure}

Fig.~\ref{fig:4a} shows the distributions of the dimensionless shear stress given by \eqref{2.17c} as $\kappa $ increases. We observe that the shear stress is nearly zero in the dislocation zone. This agrees well with the fact that the main contribution to the Peach-Koehler force comes from the shear stress, while the contribution of this force due to the boundary is noticeable only near the free boundary. Fig.~\ref{fig:5} presents on the left the normalized torque $\bar{T}$ as function of the dimensionless twist $\kappa $, where on the right the zoom of this curve near the origin is also shown. One can see that, at the onset of the dislocation nucleation (at $\kappa =\kappa _m\approx 0.000032$ for the above chosen parameters) the torque jumps down. The reason of this ``torque-drop'' is that at $\kappa =\kappa _m$ the plastic warping jumps from zero to a small but positive function leading to the reduction of the stress and also the torque. After the torque-drop there is a ``work hardening'' section followed by the softening behavior as shown on the left of Fig.~\ref{fig:5}. 

\section{Dislocation distribution at non-zero dissipation}\label{sec:distortion}

\citet{Berdichevsky2016} rightly pointed out that the solution found by the CDT in the case of zero dissipation contradicts the classical plasticity theory and to the observed behavior of real metals, where a plastic region is formed near the boundary while no plastic deformation develops in the middle of the bar. This contradiction can be resolved if we take into account the resistance to dislocation motion leading to the non-vanishing dissipation. As mentioned in Section 2, the plastic warping must then evolve in accordance with the variational equation \eqref{2.9} under the constraint $\beta (0)=0$. We regard $\omega $ as a given function of time (the ``driving'' variable) and try to determine $\beta (t,r)$. Provided the sign of $\dot{\beta }$ does not change during the evolution of $\beta $, the variational equation \eqref{2.9} reduces to minimizing the following ``relaxed energy'' functional
\begin{equation*}
I(\beta )=2\pi \int_{0}^{R} \biggl[ \frac{1}{2}\mu ( r\omega -\beta )^2  
+  \frac{\mu b}{4\pi } f(r/R) |\beta _{,r}+\beta /r|
+K\text{sign}\, \dot{\beta }\, \beta \biggr] r dr
\end{equation*}
among all admissible function $\beta (r)$ satisfying the boundary conditions $\beta (0)=0$. Finally, if $\dot{\beta}=0$, then the plastic warping is frozen, while the stress and the torque should be found with this frozen $\beta $.

Let us assume that the dislocations have the positive sign: $\beta_{,r}+\beta /r >0$.  Besides, we consider the loading process for which $\text{sign}\, \dot{\beta }=1$ so that the last term in \eqref{3.1} becomes $K \beta $. It is convenient to introduce the dimensionless quantities \eqref{2.10} and $\gamma _c=\frac{K}{\mu }$ in terms of which the functional reads 
\begin{equation}\label{3.1}
E(\beta )=\int_0^1 [\frac{1}{2}(\kappa \xi -\beta )^2+\chi f(\xi ) (\beta_{,\xi }+\beta /\xi ) +\gamma _c \beta ]\, \xi d\xi .
\end{equation}
Functional \eqref{3.1} is similar to functional \eqref{2.8}. However, the additional term $\gamma _c\beta $ changes the behavior of the solution radically. Indeed, in the case of non-zero dissipation dislocations cannot move if the shear stress is less than the critical resolved shear stress $K$. Since the stress near the middle line of the bar is always small, new dislocations cannot be formed there. Based on this deliberation, we look for the plastic warping in the form
\begin{equation}\label{3.2}
\beta (\xi )=
  \begin{cases}
  0 & \text{for $\xi \in (0,l_1)$}, \\
    \beta _1(\xi ) & \text{for $\xi \in (l_1,l_2)$}, \\
    \beta _2/\xi & \text{for $\xi \in (l_2,1)$}, 
  \end{cases}
\end{equation}
where $\beta _1(\xi )$, $l_1$, $l_2$, $\beta _2$ are unknowns. The zone $\xi \in (0,l_1)$ corresponds to the elastic core. The dislocation zone forms the ring $\xi \in (l_1,l_2)$. We admit that $\beta (\xi )$ may have jumps at $\xi =l_1$ and $\xi =l_2$. Note that this solution Ansatz is consistent with the boundary condition $\beta (0)=0$.

According to \eqref{3.2} the functional becomes
\begin{multline}\label{3.3}
E=\int_0^{l_1}\frac{1}{2}(\kappa \xi )^2\, \xi d\xi +\int_{l_1}^{l_2}[\frac{1}{2}(\kappa \xi -\beta _1)^2+\chi f(\xi ) (\beta _{1,\xi }+\frac{\beta _1}{\xi })+\gamma _c \beta _1
 ]\, \xi d\xi 
\\ 
 +\int_{l_2}^1\frac{1}{2}[(\kappa \xi -\frac{\beta _2}{\xi})^2+\gamma _c \beta _2/\xi ]\, \xi d\xi .
\end{multline}
Varying energy functional \eqref{3.3} with respect to $\beta _1(\xi)$ we obtain the (non-differential) equation for it on the interval $(l_1,l_2)$
\begin{equation}\label{3.4}
-(\kappa \xi -\beta _1)\xi -\chi f^\prime (\xi )\xi +\gamma _c \xi=0,
\end{equation}
yielding
\begin{equation}
\label{3.5}
\beta _1(\xi )=\kappa \xi -\gamma _c+\chi f^\prime (\xi ).
\end{equation}
Due to the specific linear dependence of energy functional on $\beta _{1,\xi }$ leading to the (non-differential) equation for $\beta _1(\xi)$, $\beta _1(l_1)$ and $\beta (l_2)$ cannot be varied arbitrarily at $\xi =l_1$ and $\xi =l_2$, so $\delta \beta _1(l_1)=\delta \beta _1(l_2)=0$. However, the variation of \eqref{3.3} with respect to $l_1$, $l_2$, and $\beta _0$ yields three additional conditions 
\begin{equation}\label{3.6}
\begin{split}
\frac{1}{2}\kappa ^2l_1^2-[\frac{1}{2}(\kappa l_1-\beta _1(l_1))^2+\chi f(l_1) (\beta _{1,\xi }+\frac{\beta _1}{l_1})+\gamma _c\beta _1(l_1)]=0,
\\
[\frac{1}{2}(\kappa l_2-\beta _1(l_2))^2+\chi f(l_2) (\beta _{1,\xi }+\frac{\beta _1}{l_2})+\gamma _c\beta _1(l_2)]- \frac{1}{2}(\kappa l -\frac{\beta _2}{l_2})^2-\frac{\gamma _c\beta _2}{l_2}=0, 
\\
\int_{l_2}^1 (-\kappa \xi +\frac{\beta _2}{\xi }+\gamma _c) d\xi =0.
\end{split}
\end{equation}

\begin{figure}[htb]
	\centering
	\includegraphics[width=8cm]{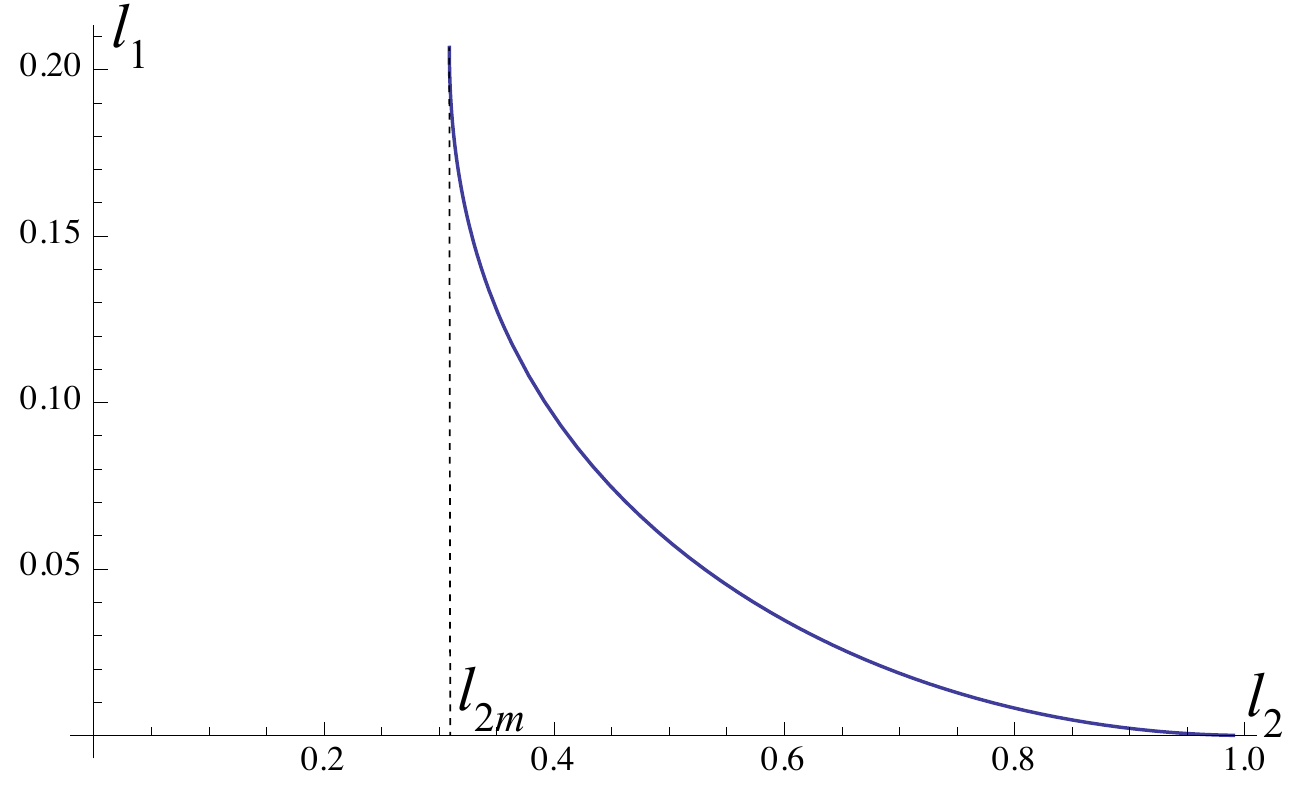}
	\caption{The plot of $l_1(l_2)$.}
	\label{fig:6}
\end{figure}

From \eqref{3.6}$_3$ we find that
\begin{equation}
\label{3.7}
\beta _2=-\frac{\kappa (1-l_2^2)-2\gamma _c(1-l_2)}{2\ln l_2}.
\end{equation}
Because $l_2<1$, $\beta _2>0$ if $\kappa >0$. Plugging \eqref{3.5} and \eqref{3.7} into \eqref{3.6}$_2$, we obtain the transcendental equation to determine $l_2$ which can be 
transformed into the quadratic equation 
\begin{equation}
\label{3.8}
a(l_2)\kappa ^2-2b(l_2)\kappa -c(l_2)=0,
\end{equation}
in terms of $\kappa $, where 
\begin{gather*}
a(l_2)=\left( l_2+\frac{1-l_2^2}{2l_2 \ln l_2}\right) ^2, \quad b(l_2)=2\chi f(l_2)+\gamma _c(l_2+\frac{1-l_2^2}{2l_2\ln l_2})(1+\frac{1-l_2}{l_2 \ln l_2}),
\\
c(l_2)=\chi ^2(f^{\prime }(l_2))^2+2\chi f(l_2)\left[ -\frac{\gamma _c}{l_2}+\chi f^{\prime \prime }(l_2)+\chi \frac{f^\prime (l_2)}{l_2}\right] -\gamma _c^2\left( 1+\frac{1-l_2}{l_2 \ln l_2}\right)^2.
\end{gather*}
We use $l_2$ as parameter and find the twist $\kappa $ through $l_2$. Provided the discriminant $b^2(l_2)+a(l_2)c(l_2)$ is positive, we take the positive root of \eqref{3.8} yielding
\begin{equation}\label{3.10}
\kappa (l_2)=\frac{b(l_2)+\sqrt{b^2(l_2)+a(l_2)c(l_2)}}{a(l_2)}.
\end{equation}
The smallest $l_2=l_{2m}$ that gives a real double root $\kappa _m=b(l_{2m})/a(l_{2m})$ is found from the equation
\begin{displaymath}
b^2(l_{2m})+a(l_{2m})c(l_{2m})=0.
\end{displaymath}
As will be seen later, this $l_{2m}$ indicates the onset of dislocation nucleation. Finally, the length $l_1$ must be found from equation \eqref{3.6}$_1$ which, after substitution of $\beta _1(l_1)$ from \eqref{3.5} and of $\kappa $ from \eqref{3.10}, reads
\begin{multline*}
\frac{1}{2}\kappa ^2l_1^2-\frac{1}{2}(\gamma _c-\chi f^\prime (l_1))^2-\chi f(l_1)[2\kappa -\frac{\gamma _c}{l_1}
\\
+\chi f^{\prime \prime }(l_1)+\chi \frac{f^\prime (l_1)}{l_1}]-\gamma _c[\kappa l_1-\gamma _c+\chi f^\prime (l_1)]=0.
\end{multline*}
Solving this equation, we find $l_1$ as function of $l_2$. The plot of this function for $l_2\in (l_{2m},1)$ is shown in Fig.~\ref{fig:6}. We see that $l_1(l_2)$ is a monotonically decreasing function. At the onset of dislocation nucleation (at $l_2=l_{2m}$) $l_1$ achieves the maximum which is not much less than $l_2$. This means that the dislocation zone at the onset of the dislocation nucleation is a thin ring. As $l_2$ (and $\kappa $) increases, $l_1$ decreases, so the ring occupied by dislocations expands during the loading process while the elastic core region diminishes. 

Having $\beta (\xi )$ according to \eqref{3.2} and \eqref{3.5}, we find the signed dislocation density from \eqref{2.2a}. For $\xi \in (0,l_1)$ and $\xi \in (l_2,1)$ we have $\rho =0$. For $\xi \in (l_1,l_2)$
\begin{equation}
\label{3.11}
\rho =\frac{1}{b}(\beta _{,r}+\frac{\beta }{r})=\frac{1}{bR}(\beta _{,\xi }+\frac{\beta }{\xi})=\frac{1}{bR}\left[ 2\kappa -\frac{\gamma _c}{\xi }+\chi (f^{\prime \prime }(\xi)+\frac{f^\prime (\xi)}{\xi}) \right] .
\end{equation}
The dimensionless shear stress distribution $\tau (\xi )=\sigma _{\theta z}(\xi )/\mu $ is given by
\begin{equation}
\label{3.13}
\tau (\xi )=\begin{cases}
\kappa \xi & \text{for $\xi \in (0,l_1)$}, \\
\gamma _c-\chi f^\prime (\xi )   & \text{for $\xi \in (l_1,l_2)$}, \\
  \kappa \xi + \frac{\kappa (1-l_2^2)-2\gamma _c(1-l_2^2}{2\ln l_2 \, \xi }  & \text{for $\xi \in (l_2,1)$}.
\end{cases}
\end{equation}
In the elastic zone the stress obeying Hooke's law is a linear function of $\xi $. In the zone occupied by dislocations, due to the smallness of $\chi $, the stress is nearly constant and equals the critical resolved shear stress. The torque is computed as the derivative of the functional \eqref{3.1} with respect to $\kappa $ giving
\begin{equation*}
\bar{T}=\frac{\partial E}{\partial \kappa}=\int_0^1 (\kappa \xi -\beta )\xi ^2d\xi .
\end{equation*}
Substituting $\beta (\xi )$ from \eqref{3.2} into this integral, we easily find that
\begin{multline}
\label{3.14}
\bar{T}=\frac{\kappa }{4}l_1^4+\frac{1}{3}\gamma_c(l_2^3-l_1^3)-\chi (l_2^2-l_1^2+\ln \frac{1-l_2^2}{1-l_1^2})
\\
+\frac{\kappa }{4}(1-l_2^4)+\frac{\kappa (1-l_2^2)^2-2\gamma _c(1-l_2)}{4\ln l_2}(1-l_2^2).
\end{multline} 
For the twist $\kappa <\kappa _m$ the plastic warping $\beta (\xi )$ must be identically zero, so we have the purely elastic solution with $u_\theta (\xi )=\kappa \xi $ and $\beta (\xi )=0$. In this case the torque $\bar{T}$ is proportional to the twist $\kappa $: 
\begin{displaymath}
\bar{T}=\frac{\kappa }{4}.
\end{displaymath}

\begin{figure}[htb]
	\centering
	\includegraphics[width=8cm]{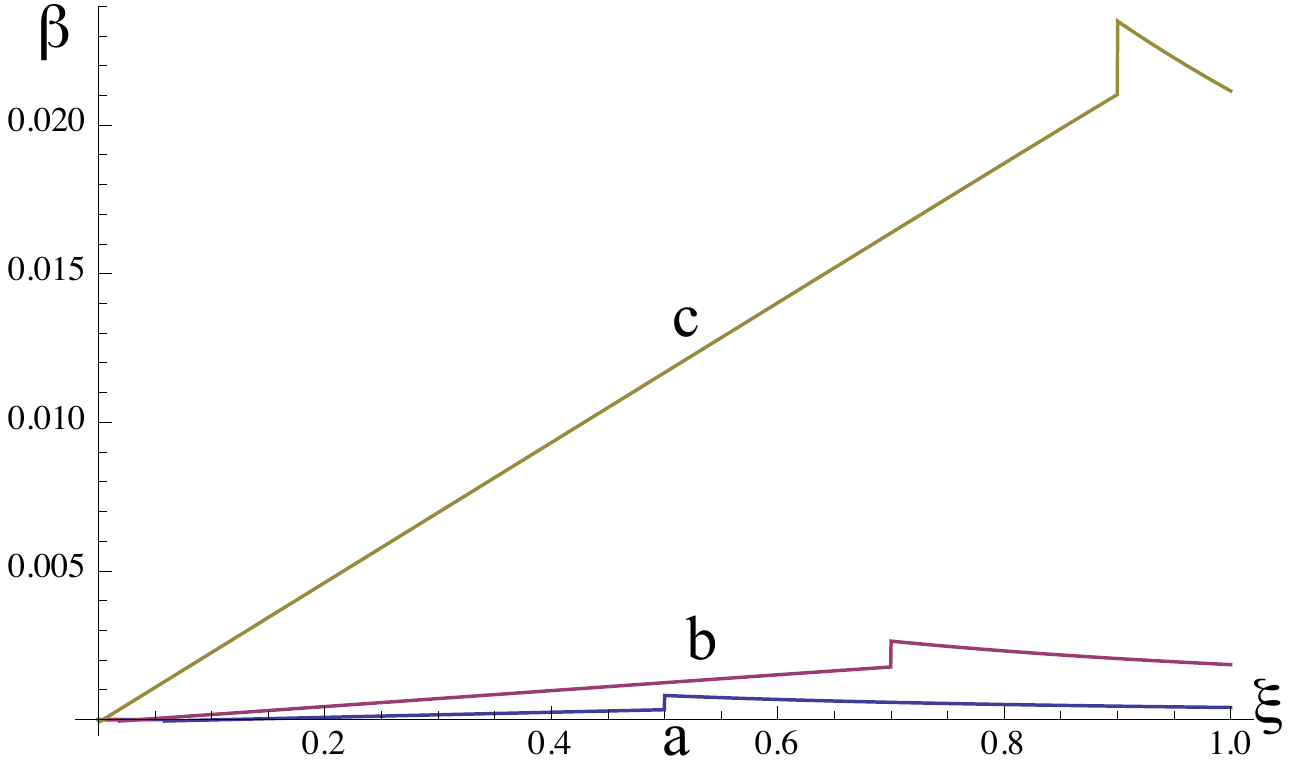}
	\caption{Evolution of the plastic warping $\beta (\xi )$: (a) $l=0.5$ ($\kappa =0.00088$), (b) $l=0.7$ ($\kappa =0.0027$), (c) $l=0.9$ ($\kappa =0.02356$).}
	\label{fig:7}
\end{figure}

Fig.~\ref{fig:3} shows the evolution of $\beta (\xi )$ as $\kappa $ increases. For the simulation we took $R=1$micron, $b=r_0=1$\AA , and $\gamma _c=10^{-4}$. As in the previous case, $\beta (\xi )$ increases with $\kappa $. Besides, the plastic warping exhibits jumps at $\xi =l_1$ and $\xi =l_2$. The difference to the case of zero dissipation is that there is a an elastic zone with $\beta =0$. The radius of the elastic zone diminishes with the increasing twist.

\begin{figure}[htb]
	\centering
	\includegraphics[width=8cm]{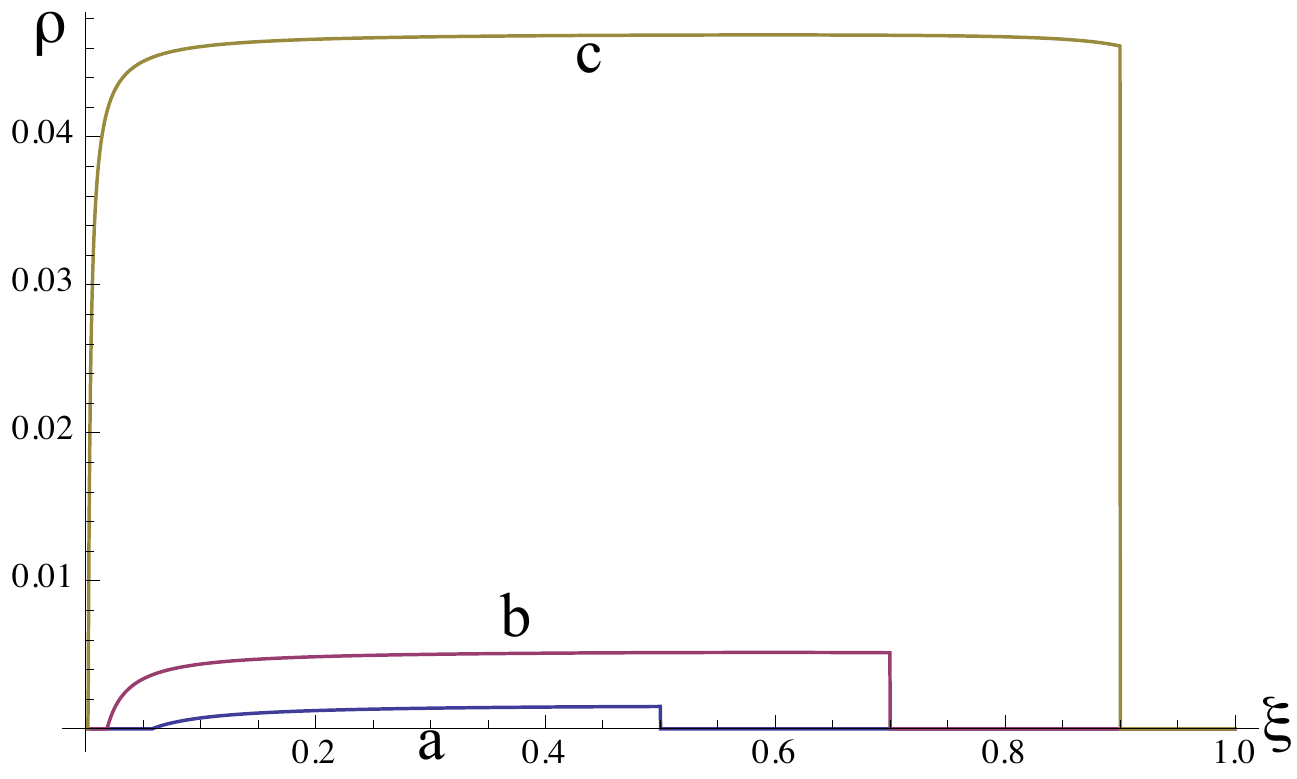}
	\caption{Distribution of normalized dislocation density $\bar{\rho }(\xi )$: (a) $l=0.5$ ($\kappa =0.00088$), (b) $l=0.7$ ($\kappa =0.0027$), (c) $l=0.9$ ($\kappa =0.02356$).}
	\label{fig:8}
\end{figure}

Fig.~\ref{fig:8} shows the distributions of the normalized dislocation density $\bar{\rho }(\xi )=bR \rho (\xi )$ as $\kappa $ increases. One can see that the dislocation-free zones at the origin as well as near the free boundary diminish as $\kappa $ increases. For small $\kappa $ the dislocation density remains nearly constant except near the origin. For large $\kappa $ we see some influence of the free boundary on the distribution of dislocations: the density is slightly decreases with $\xi $. Note that the dislocation density is continuous at $\xi =l_1$ and discontinuous at $\xi =l_2$.

\begin{figure}[htb]
	\centering
	\includegraphics[width=8cm]{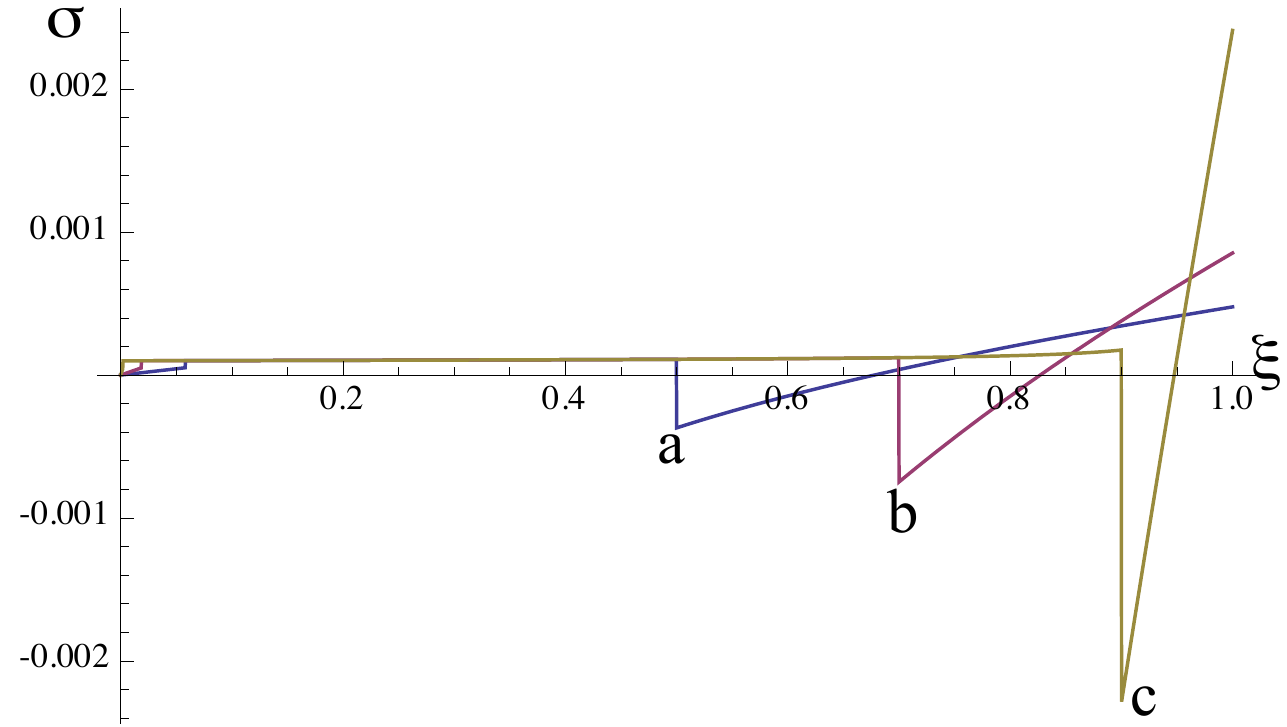}
	\caption{Distribution of dimensionless shear stress $\tau (\xi )$: (a) $l=0.5$ ($\kappa =0.00088$), (b) $l=0.7$ ($\kappa =0.0027$), (c) $l=0.9$ ($\kappa =0.02356$).}
	\label{fig:9}
\end{figure}

\begin{figure}[htb]
	\centering
	\includegraphics[width=13cm]{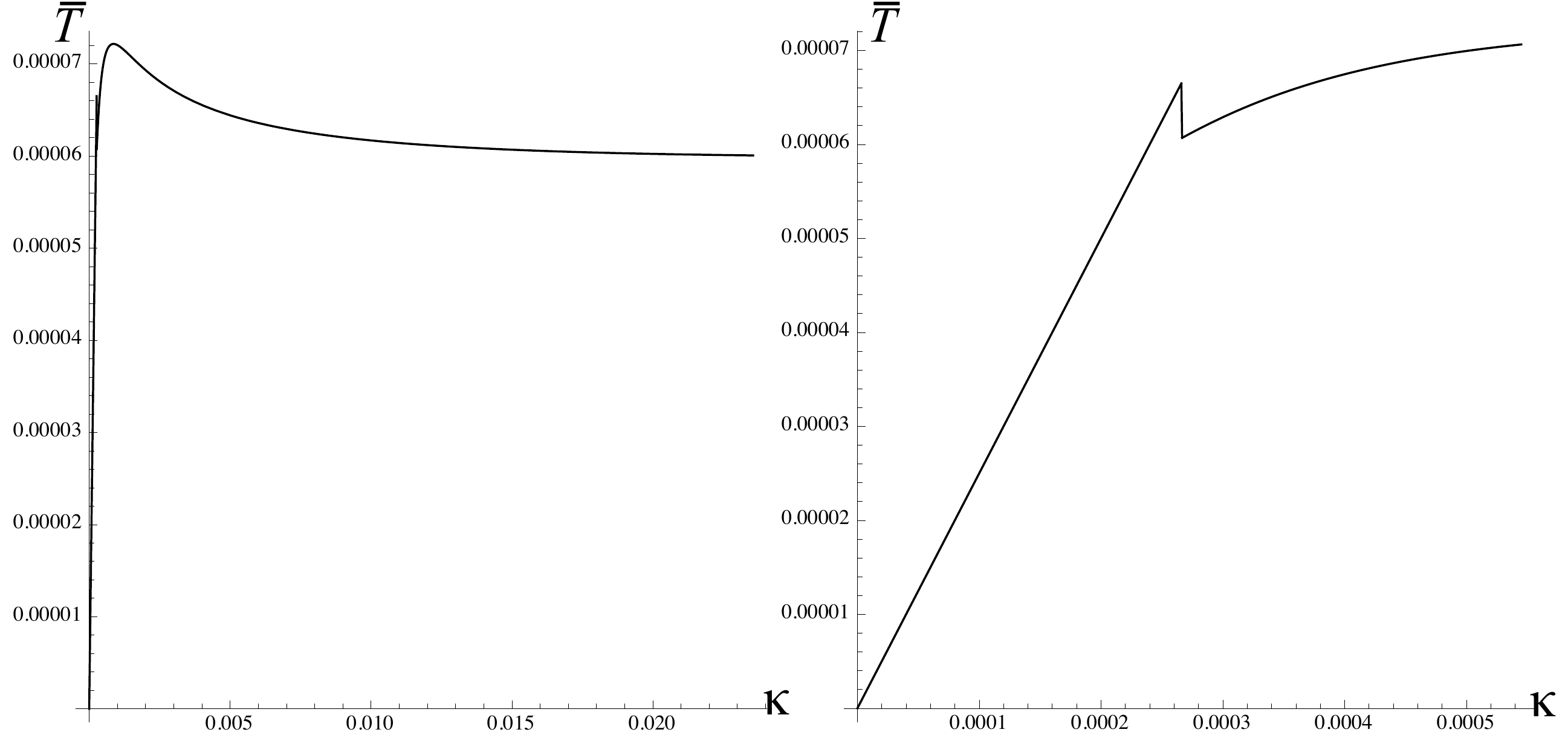}
	\caption{The torque-twist curve: (i) Left: the whole curve, (ii) Right: Zoomed near the origin.}
	\label{fig:10}
\end{figure}

Fig.~\ref{fig:9} shows the distributions of the dimensionless shear stress given by \eqref{2.17c} as $\kappa $ increases. We observe that the shear stress increases first as a linear function in the elastic zone, then remain nearly constant (which is equal to $K$) in the dislocation zone, and finally jumps down and increases linearly in the outer dislocation-free ring. Fig.~\ref{fig:10} presents on the left the normalized torque $\bar{T}$ as function of the dimensionless twist $\kappa $, where on the right the zoom of this curve near the origin is also shown. One can see that, at the onset of the collective dislocation nucleation (at $\kappa =\kappa _m\approx 0.00027$ for the above chosen parameters) the torque jumps down. The reason of this ``torque-drop'' is that at $\kappa =\kappa _m$ the plastic warping jumps from zero to a small but positive function leading to the reduction of the stress and also the torque. After the torque-drop there is a ``work hardening'' section followed by the softening behavior as shown on the left of Fig.~\ref{fig:10}. Note that the softening effect is much less pronounced as in the case without dissipation.  

\section{Conclusion}
In this paper we have shown that the torsion of the bar of circular cross section can be analytically solved within the asymptotically exact continuum theory of dislocations. If the resistance to dislocation motion is negligible, then dislocations are concentrated in a circle of radius less than the radius of the cross-section. The outer ring is dislocation-free. The plastic warping suffers a jump across the boundary between dislocation and dislocation-free regions indicating misorientation of the crystal lattice. There is a threshold torque for dislocation nucleation. The torque drop takes place at the onsets of dislocation nucleation followed by the short hardening and subsequent softening. If the resistance to dislocation motion is taken into account, then there exist the elastic zone in the middle of the cross-section. Dislocations are concentrated in a ring, whose size increases as the twist increases. The torque drop is also observed at the onsets of dislocation nucleation, but the hardening and softening effects are much less pronounced compared to the case without dissipation.

Last but not least, it would be quite convincing if this theoretical result for the dislocation distribution could be compared with the experimental observations and measurements. It is hoped that this paper would serve as motivation for experimentalists using EBSD-technique (see, e.g., \citep{Kysar2010}) or, alternatively, the etch pits method to observe the dislocation zone in twisted bars and to measure the dislocation density.

\end{document}